\documentclass[review]{elsarticle}

\usepackage{lineno,hyperref}
\modulolinenumbers[5]

\journal{Journal of \LaTeX\ Templates}
\usepackage{xspace}
\usepackage{pdfwidgets}
\usepackage{enumitem}
\usepackage{latexsym}
\usepackage{tikz}
\usepackage{pgfplots}
\usepackage{natbib}
\pgfplotsset{compat=newest}
\usepgfplotslibrary{groupplots}
\usetikzlibrary{pgfplots.groupplots}
\usepgfplotslibrary{colormaps}
\usetikzlibrary{pgfplots.colormaps}
\usetikzlibrary{calc}
\usepgfplotslibrary{external}
\tikzsetexternalprefix{figures/}
\usepackage[textsize=tiny]{todonotes}
\makeatletter
\renewcommand{\todo}[2][]{\tikzexternaldisable\@todo[#1]{#2}\tikzexternalenable}
\makeatother

\makeatletter
\def\bs{\expandafter\@gobble\string\\}
\def\lb{\expandafter\@gobble\string\{}
\def\rb{\expandafter\@gobble\string\}}
\def\@pdfauthor{C.V.Radhakrishnan}
\def\@pdftitle{elsarticle.cls -- A documentation}
\def\@pdfsubject{Document formatting with elsarticle.cls}
\def\@pdfkeywords{LaTeX, Elsevier Ltd, document class}

\DeclareRobustCommand{\LaTeX}{L\kern-.26em%
        {\sbox\z@ T%
         \vbox to\ht\z@{\hbox{\check@mathfonts
           \fontsize\sf@size\z@
           \math@fontsfalse\selectfont
          A\,}%
         \vss}%
        }%
     \kern-.15em%
    \TeX}
\makeatother

\usepackage{array}
\newcolumntype{P}[1]{>{\centering\arraybackslash}p{#1}}

\begin{document}

\begin{frontmatter}

\title{Analysis and Suppression of RF Radiation from the PSI 590MeV Cyclotron Flat Top Cavity}


\author[PSI]{N. J. Pogue\corref{mycorrespondingauthor}}
\ead{nathaniel.pogue@psi.ch}
\ead[url]{psi.ch}

\address[PSI]{Paul Scherrer Institut, Villigen-PSI, Switzerland}



\cortext[mycorrespondingauthor]{Corresponding author}


\author[PSI]{A. Adelmann\corref{mycorrespondingauthor}}
\ead{andreas.adelmann@psi.ch}
\author[PSI]{L. Stingelin\corref{mycorrespondingauthor}}
\ead{lukas.stingelin@psi.ch}

\begin{abstract}
The Flat Top Cavity, located in the PSI HIPA Ring Cyclotron leaks RF power of several kilo Watts into the cyclotrons vacuum space causing several complications. A detailed electromagnetic model was created and simulations performed to analyze the mechanisms by which power is leaking out of the Flat Top Cavity. The tolerances needed to limit the leaked power in future iterations of the Flat Top cavity are reported. Comparison of the model to measurements are described as well as two potential methods to limit power leakage. These studies will have direct impact on future RF cavity designs for cyclotrons as power levels increase and higher RF fields are required.
\end{abstract}

\begin{keyword}
\texttt{Flat Top \sep RF Cavity \sep Cyclotron \sep Multi-physics \sep Simulation}
\MSC[2010] 00-01\sep  99-00
\end{keyword}

\end{frontmatter}

%
%
%

\section{Introduction}

The Flat Top Cavity (FTC) in the PSI HIPA Ring Cyclotron \cite{Ring} is one of the critical devices required to achieve a continuous 590~ MeV, 2.2~mA proton beam. Operating in the 3rd harmonic, the fields have an effect of increasing the longitudinal acceptance of the machine. The addition of a slight slope to the voltage confines the beam and reduces longitudinal emittance growth and compensates for space charge effects \cite{Joho}. 

\begin{figure}[h!]
\centering
   \includegraphics[scale=.5]{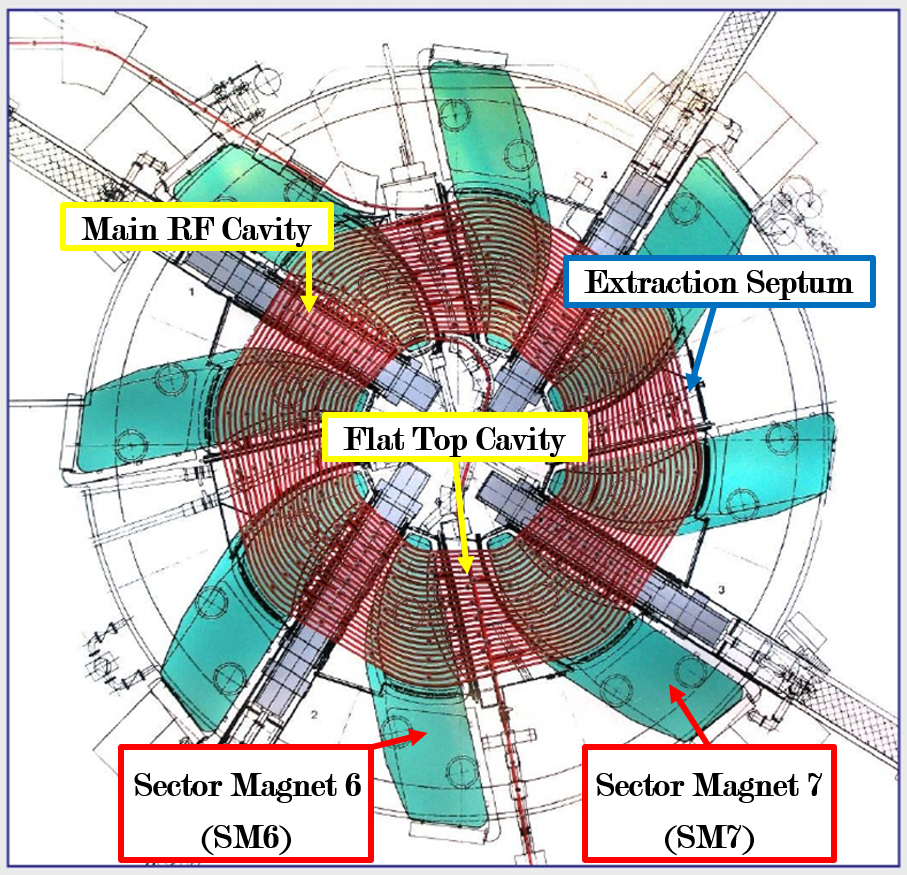}
  \caption{Image shows a schematic of the PSI Ring Cyclotron highlighting the Flat Top Cavity and the two sector magnets that flank it.}
  \label{Figure1}
\end{figure}

The PSI Ring cyclotron, Figure \ref{Figure1} obtained its world record holding current using this scheme over the past 41 years through incremental upgrades to the entire system. Originally designed for 100 $\mu$A, the cyclotron has evolved greatly and with it the proton beam intensity has dramatically increased. Most of the system have been upgraded, including the four main RF cavities \cite{Fitze}. The FTC, which up until January 2015 had not been moved in 35 years, is one of the oldest components in the system. It is also the piece of hardware preventing the machine reaching higher currents.

The FTC$'$s limitation is a complex problem. The power being inserted into the cavity causes the surface to move and deform from predominantly thermal expansion caused by the power dissipation on the aluminum surface. This results in cavity detuning. Hydraulic systems are in place to correct for this variation but they have reached their maximum pressure capacity. Additionally the amount of power needed to reach the 560~kV peak voltage (without beam) is near the limitations of the RF amplifier and coaxial-line. Lastly, though not a limitation to the cavity itself, a plasma is being generated by the FTC within the vacuum space located within the sector magnets. This plasma may be causing the metal deposition observed on the critical ceramic components as well as coupling power out of the cavity \cite{Plasma}.

The combination of these problems, and the cavity$'$s age, has led to investigations focused on: the current status of the cavity, the existing RF model for the cavity, minimally invasive interventions that could be performed, and finally a new cavity design that could provide sufficient voltage to allow the main cavities to achieve peak operation. Below is a description of the models and experimental data used to understand the cavity$'$s operation and establishment of parameters that must be adhered to for future iterations of the FTC.

\section{RF Models}

 \begin{figure}[h!]
\centering
   \includegraphics[scale=1.2]{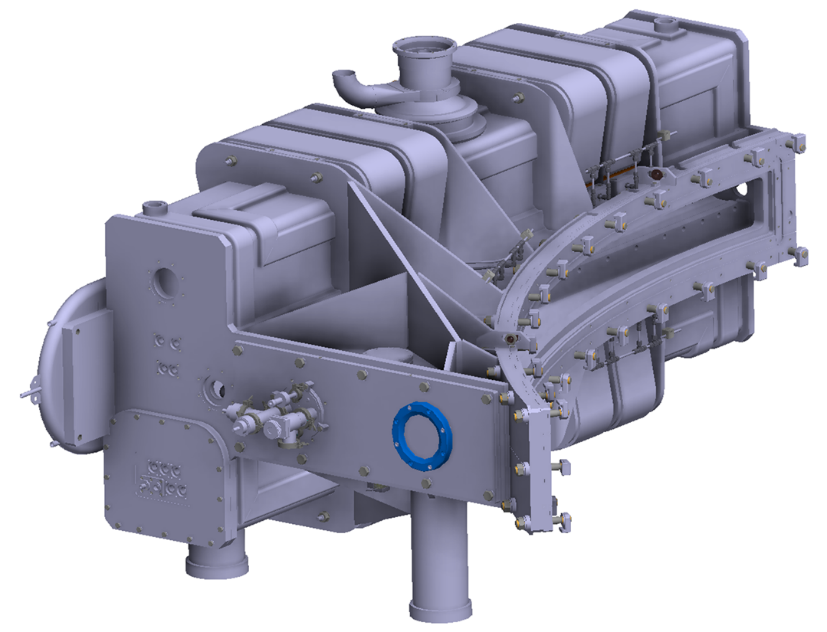}
  \caption{The 3D CAD model of the Flat Top Cavity}
  \label{Figure2}
\end{figure}

 \begin{figure}[h!]
\centering
   \includegraphics[scale=.5]{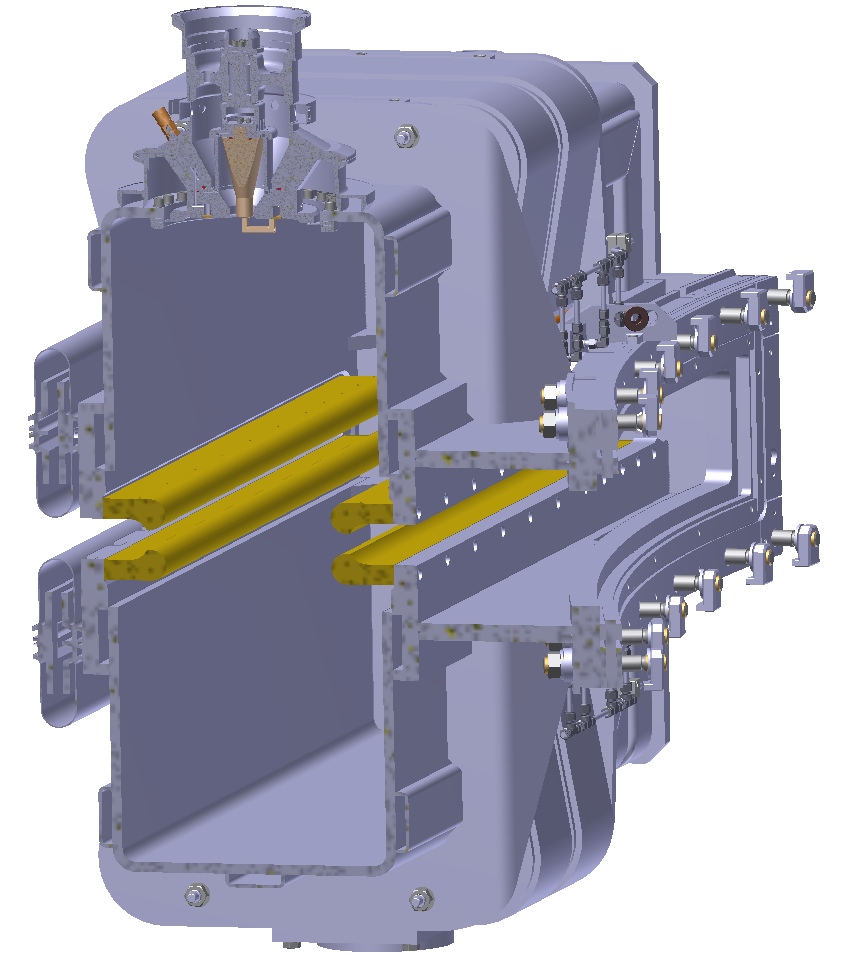}
  \caption{A cross section of the cavity is presented highlighting the lips in gold. The lips$'$ locations are critical to reduce leaking power. Additionally the coupler is shown in copper.}
  \label{CavCross}

\end{figure}

 \begin{figure}[h!]
\centering
   \includegraphics[scale=.33]{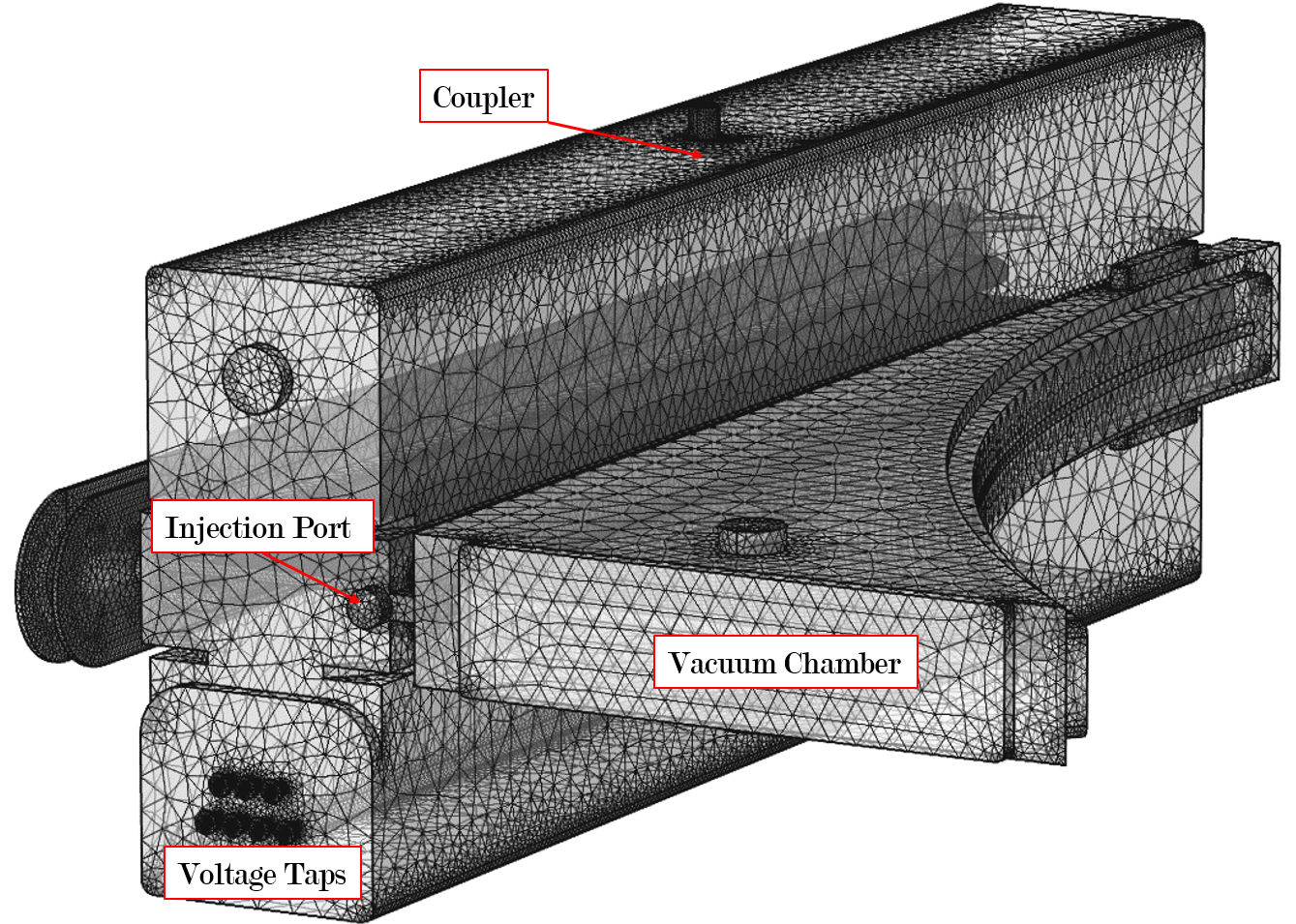}
  \caption{The model created to be used in the tolerance and power loss simulations. The new model takes into account the vacuum space between the cavity and Sector Magnet 7 (SM7), dielectric covers of the voltage taps, and several ports - most notably the injection beam line ports.}
  \label{Figure3}

\end{figure}

A completely new and detailed model was made from the 3D CAD drawings (shown in Figures \ref{Figure2} and \ref{CavCross}). It has the capability to be parameterized for easy manipulation of various features. The empty space within the 3D cavity was extracted and several ports were added. Significant features to note are: the different size and shape for each of the two apertures leading out of the cavity to the sector magnets, the manhole-flange which allows entry into the cavity, the entire vacuum chamber between the cavity and the Sector Magnet 7 (SM7), and lastly an Aquadag \cite{Aqua} thin film coating was placed on the corresponding surfaces with the adjusted surface resistance.

\begin{figure}[h]
 \centering
\hspace*{-0cm}
   \includegraphics[scale=.84]{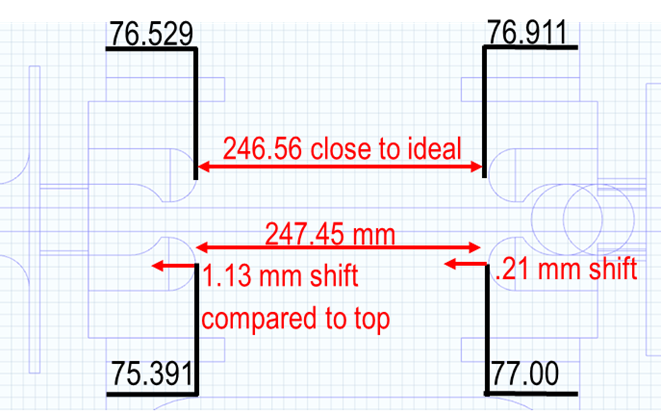}
  \caption{Location of the lips reference against the average of the side walls. The separation and position on top is close to the ideal design, however the bottom is laterally shifted and very asymmetric.}
  \label{Figure4}
\end{figure}

The resulting model is shown in Figure \ref{Figure3} and has excellent agreement with the existing cavity$'$s unloaded quality factor, Q$_{0}$. The Q$_{0}$ of the cavity was measured to be 28,287$\pm$58 and the Q$_{0}$ of the modeled cavity is 28,213. Using laser interferometry, the cavity$'$s inner surface was mapped to a 7-8~ppm accuracy in distance between the surface and the tracker. These distance were then calculated and a 3D map of the cavity is generated from the data. The results confirmed the general dimensions of the CAD model, but also showed some irregularities. The first instance was that the lips were not vertically aligned on their respective sides, but were aligned horizontally, see Figure \ref{Figure4}. On each side, a difference of 1-2~mm occurred between the top and bottom lips. This was initially thought to be rather inconsequential as 2~mm difference in a cavity that is 2.7~m long is rather trivial. However this was found to be an incorrect assumption.

 \begin{figure}[h!]
 \centering
\hspace*{-0cm}
   \includegraphics[scale=1.2]{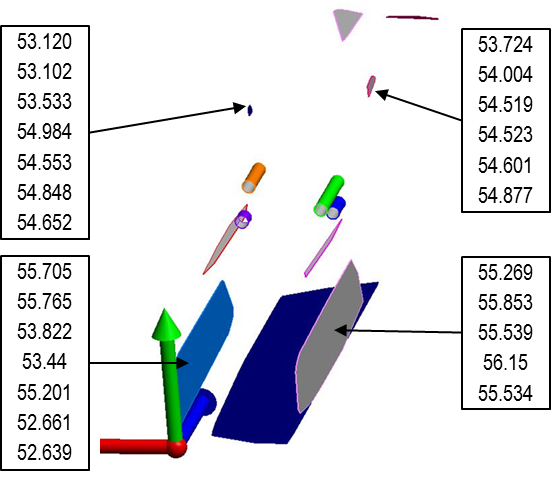}
  \caption{The image is of the interferometry measurements performed. The surfaces generated show the position of the cavities walls and lips. For the simulation, points along the surfaces were measured and quantified to establish and average values to be used for the four walls and lips in the model.}
  \label{inter}
\end{figure}

Observing the data closely, the entire cavity is warped and varies along its height and length as illustrated in Figure \ref{inter}. In order to model the cavity with full accuracy, the entire interior profile would have to be measured, which is unrealistic for several reasons (the most important being that the cavity is radioactive).

Even though the entire cavity was not mapped, several interesting features were revealed. First, the top half of the cavity is shifted horizontally from the bottom half. Second, the cavity gets wider as you move towards the center. Lastly, the cavity grows larger as you move from SM6 side to SM7. The walls of the cavity are also not planar. The walls bow and twist which can be seen from measured points around the surface. As a result, to determine an ideal case and a realistic case, the averages of the walls$'$ positions were used.

Due to the variations throughout the cavity, a parametric study was performed to determine the positioning tolerances on the ``lips", the most critical feature in the cavity. From Figure \ref{Figure4} it can be seen that each lip protrudes into the cavity at a different length. To create an ideal case, the lips were perfectly aligned and inserted into the CAD based RF model. The ``ideal case" was altered by moving the position of each lip incrementally and determining the effect on the cavity. A second case was also developed which used the average measured position of the lips along the cavity length. This model will be referred to as the "realistic case".

The fine placement of the lips will be shown to be the dominant term in the amount of power leaking out of the cavity in to the vacuum space. In turn, the leaked power is igniting a plasma and creating secondary emission of particles in various regions in the cyclotron vacuum space \cite{Plasma}. Through these studies it is hoped to identify the tolerance needed to minimize leaked power and identify methods to improve the existing cavity, as well as establish criteria needed for future cyclotron cavities.

\section{Tolerance Study}

For RF cavities, symmetry is sought to reduce variation in field profiles and reduction of higher order modes. In the case of the FTC there is also a direct correspondence between asymmetry and the amount of power leaked out of the cavity. The position of the lips is the most influential feature in the entire cavity with regards to voltage gain, frequency, peak electric field, and power leakage. The initial assumption was that 1-2~mm was not a detrimental position change in a cavity that is 2.7~m long, 0.4~m wide, and 1.5~m tall. This was an incorrect assumption, as the initial studies into the cavity illustrated, see Table \ref{table1}. For reference, the SM7 side is the side where the vacuum chamber is located between the FTC and the sector magnet.

\begin{figure}[h]
 \centering
\hspace*{-0cm}
   \includegraphics[scale=.33]{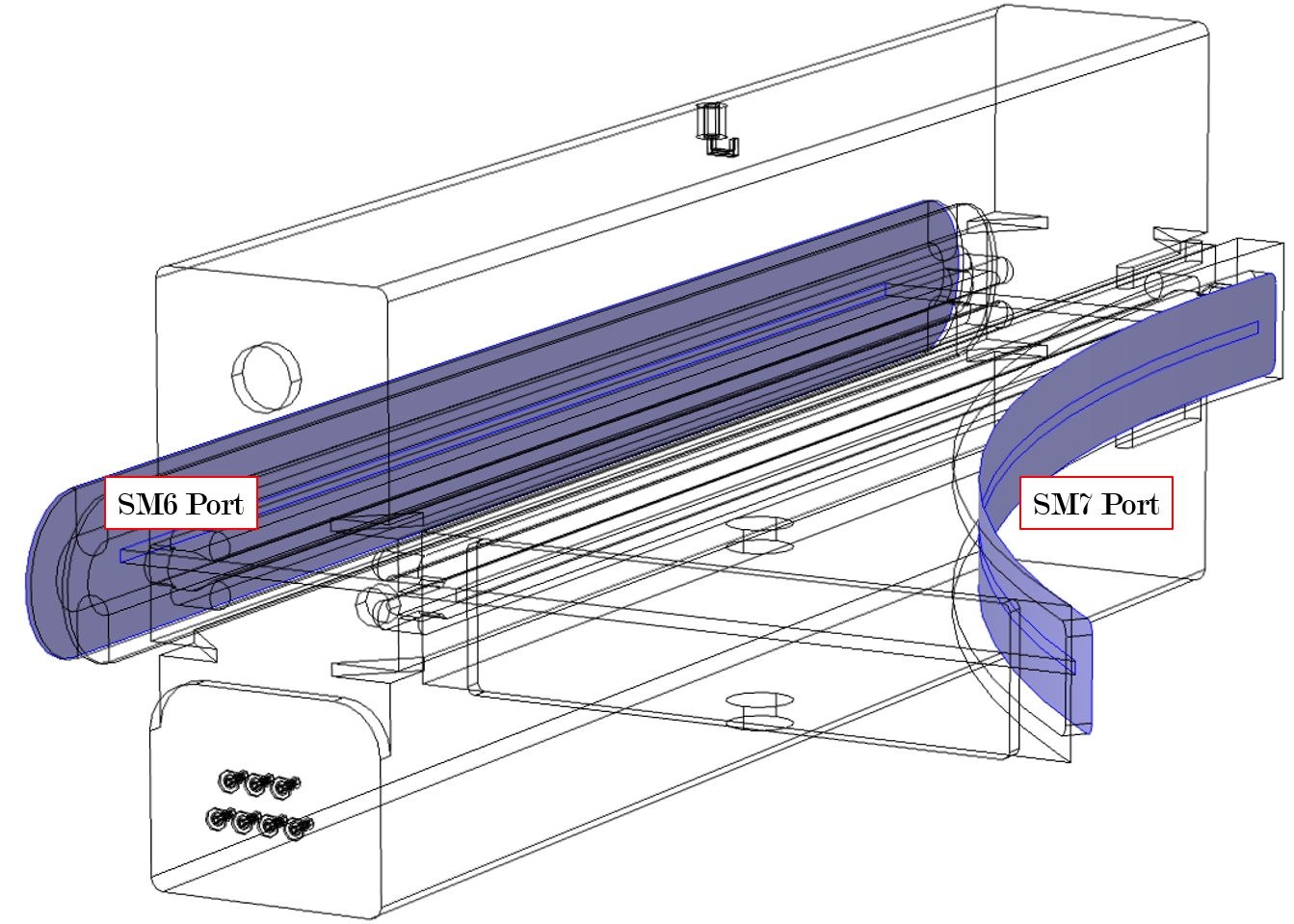}
  \caption{Image illustrates the locations where power leakage was measured on the SM6 (left) and SM7 (right) sides.}
  \label{Figure6}
\end{figure}

The study was performed with COMSOL Multiphysics \citep{Comsol}, specifically the Eigenvalue and Frequency Domain solvers. Impedance boundary conditions were placed on the walls, taking the physical characteristics of the material in the cavity into account. For determining the power leaked, scattering boundary conditions were placed on the end ports making the wall transparent to the electromagnetic fields. Approximately 1.3 million tetrahedral elements were used during the simulation.

\begin{table}[h!]
 \centering
\hspace*{-0cm}

\begin{tabular}{ |P{2.5cm}||P{1.5cm}|P{1.5cm}|P{1.5cm}|  }
 \hline
 \multicolumn{1}{|c||}{}&\multicolumn{2}{c|}{Leaking Power} &\multicolumn{1}{|c|}{Voltage}  \\
 \hline
 Case& SM6 (W)&SM7 (W)& (kV)\\
 \hline
 Ideal   & 271    &334&  496\\
 1 mm Top SM6&   1735  & 355   &475\\
 1 mm Top SM7& 233 & 2565&  453\\
 Positions from	 Interferometry& 468 & 7974&  411\\
 \hline
\end{tabular}
  \caption{Summary of initial studies into the leaked power of the FTC. The power inserted into the cavity was set to reach 500 kV with leakage not allowed. Then ports were opened allowing RF to escape. In the ideal case, or perfectly symmetric case, very little energy was lost. The voltage dropped by only 4 kV. However, with larger perturbations to the lips position the amount of energy lost increases dramatically and drops the achievable voltage. The worst case was the lips in the position dictated by the interferometry measurements of the physical cavity.}
  \label{table1}
  \end{table}

The ``ideal case", when the lips are vertically aligned, causes a few Watts of power to be leaked out through the beam apertures. The study was scaled to insert sufficient power to create a peak voltage gain of 500~kV when power is not allowed to leak out of the system. The ``ideal case" dropped to 496~kV due to the $\sim$600 W leaked out of the cavity. The lips$'$ positions were then varied by only 1~mm on each side independently. When the lips were placed in an asymmetric position, the amount of power leaked out of the cavity became large, causing the voltage to drastically drop. The effect was greatly amplified when the lip was moved 2~mm off alignment instead of 1~mm shown above. The positions of the lips according to the interferometry, or ``realistic case", were used in the model and led to nearly 8000 W~of leaked power.

\begin{table}[h!]
 \centering
\hspace*{-0cm}

\begin{tabular}{ |P{1cm}||P{1cm}|P{1cm}|P{1cm}|P{1cm}||P{1cm}|P{1cm}|P{1cm}|P{1cm}|P{1cm}| }
 \hline
 \multicolumn{1}{|c||}{}&\multicolumn{4}{c||}{SM7 Top (kW)}&\multicolumn{4}{c|}{SM7 Bottom (kW)} \\
 \hline
 Shift& Total&Leaked&SM7 Port& SM6 Port& Total&Leaked&SM7 Port& SM6 Port\\
 \hline
 0.0		&73.9		&0.63		&0.35		&0.28		&&&&\\
 0.5		&72.6		&1.34		&1.12		&0.22		&72.5&0.91&0.75&0.17\\
 1.0		&74.6		&3.63		&3.36		&0.27		&74.1&2.87&2.70&0.17\\
 1.5		&78.0		&7.28		&6.96		&0.32		&77.1&6.07&5.91&0.17\\
 2.0		&82.7		&12.3		&11.9		&0.39		&81.5&10.82&10.63&0.19\\
 \hline
 \multicolumn{1}{|c||}{}&\multicolumn{4}{c||}{SM6 Top (kW)}&\multicolumn{4}{c|}{SM6 Bottom (kW)} \\
 \hline
 Shift& Total&Leaked&SM7 Port & SM6 Port& Total&Leaked&SM7 Port& SM6 Port\\
 \hline
 0.0		&73.9		&0.63		&0.35		&0.28		&&&&\\
 0.5		&72.6		&1.19		&0.26		&0.93		&72.1&0.65&0.24&0.41\\
 1.0		&74.1		&2.90		&0.29		&2.61		&73.2&1.88&0.26&1.62\\
 
 1.5		&76.6		&5.56		&0.31		&5.25		&75.1&3.99&0.27&3.72\\
 2.0		&80.0		&9.18		&0.36		&8.82		&78.1&7.19&0.32&6.87\\
 \hline
\end{tabular}

   \caption{The chart shows the total power required by the cavity for the specified conditions of asymmetry. The total leaked power is shown as well as which side of the cavity the power is entering the cyclotrons vacuum space. Each row shows the dependence of the leaked power of each lip moved independently of the others in 0.5 mm increments outward from the center of the cavity. All the lips, excluding the one moved, are in their initial ``Ideal Case" positions. Note that the 0 shift case is the ideal case where the lips are all symmetrically aligned for reference. The top lips$'$ positions are more sensitive than the bottom lips$'$ position. Secondly the SM7 side is more sensitive to perturbation.}
  \label{table2}
  \end{table}

These initial studies, which show slight deviations can cause staggering losses, led to re-evaluation of the position of the lips with respect to each other and their location in reference to the walls. For the ``ideal case", the positions of the lips were made symmetric, vertically and horizontally, in the cavity and separated by 246.56~mm, which is the separation of the top lips of the cavity as shown in Figure \ref{Figure4}. Then each one of the four lips was moved independently in 0.5~mm increments. The results of the power leaked out of the beam ports (Figure~\ref{Figure6}) is tabulated in Table \ref{table2}.

In the tolerance study, the cavity was set to always reach 500~kV peak voltage, regardless of the amount of power leaked out from the boundary. Compared to previous studies, the real coupler shape was inserted into the cavity. Finer mesh was placed on the lips, dielectric materials were inserted in the cavity, and quantitative calculation of the Aquadag resistance on the aluminum were all inserted into the cavity$'$s parameters. The ``ideal case" has very little leakage with respect to the amount of power needed to get the aluminum cavity up to field ($\sim$71~kW). As the lips are moved one at a time in 0.5~mm steps, the power leakage increase dramatically.

Several interesting features were observed in this study. The power always leaks predominantly out of the side with the introduced asymmetry. A slight asymmetry can cause a slight decrease in leaked power on the opposite side of the asymmetrical feature. Lastly, the top half of the cavity is more sensitive to perturbation than the bottom half. This is most likely due to three features of the cavity. First, the coupler (located in the top half) is not positioned centrally but rather shifted towards SM7. The other two features are the dielectric covers for the pickups and the added volume from the manhole entrance on the lower half creates a higher stored energy in the bottom half of the cavity.

Another interesting result was that the SM7 side leaks more than the SM6 side. This difference is believed to be caused by the fact that the beam pipe outlet from the cavity is larger on the SM7 side (0.161 m$^2$) compared to the SM6 side (0.155 m$^2$). The position of the coupler may also contribute to this difference between the two sides but is limited to a maximum of 12\% of the additional leaked power.

 \begin{figure}[h!]
 \centering
\hspace*{-0cm}
   \includegraphics[scale=.31]{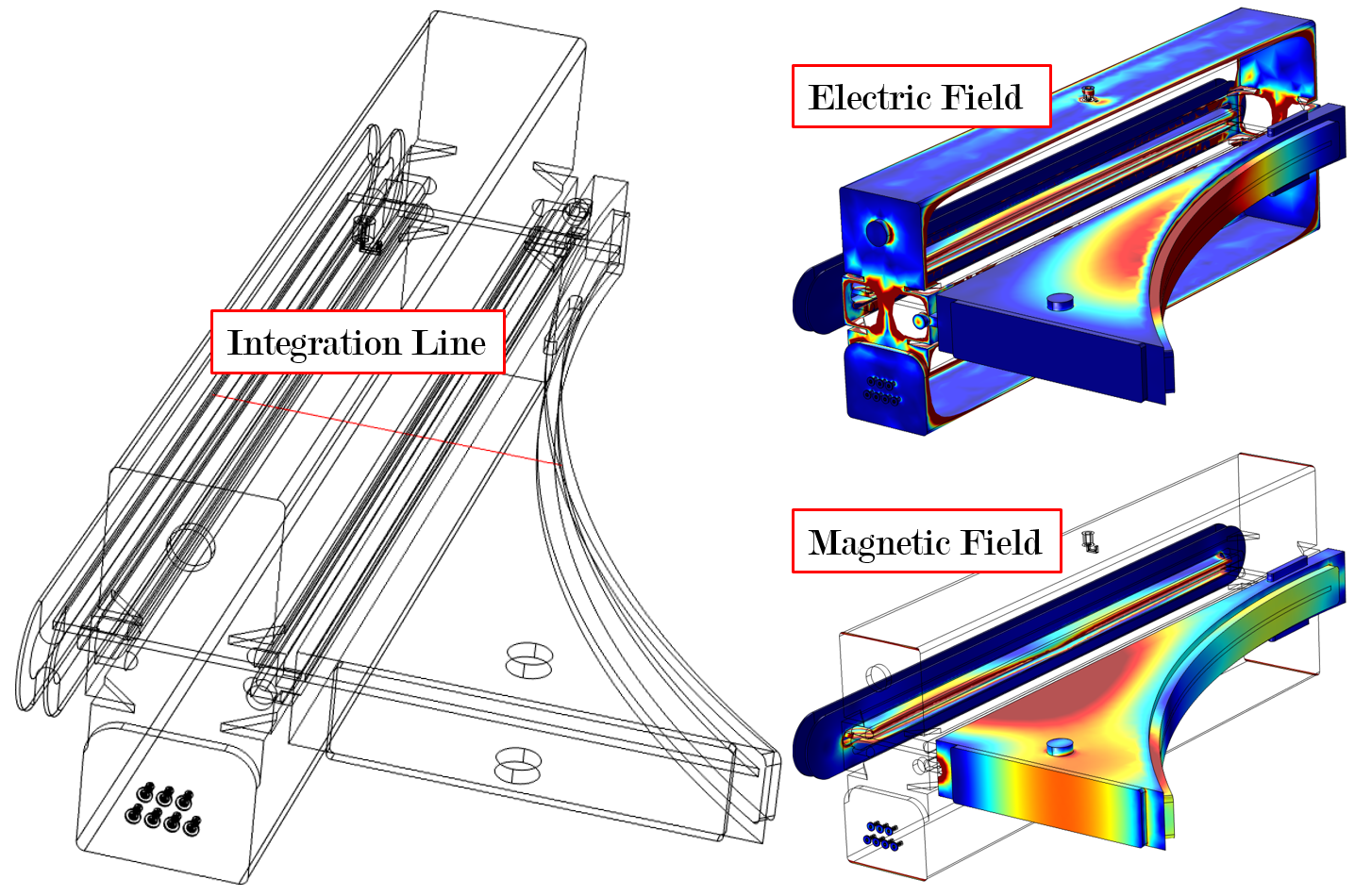}
  \caption{(Top) The "realistic case" model and the line integration used to determine the voltage gain in the cavity. The electric (top right) and the magnetic (bottom right) field patterns are shown. The integration through the vacuum chamber adds $\sim$.5$\%$ to the voltage and $\sim$1.2$\%$ reduction on power needed to reach the field in the Realistic case. The field pattern changes depending on the lip$'$s position.}
  \label{vacgain}
\end{figure}

 \begin{figure}[h!]
 \centering
\hspace*{-0cm}
   \includegraphics[scale=.33]{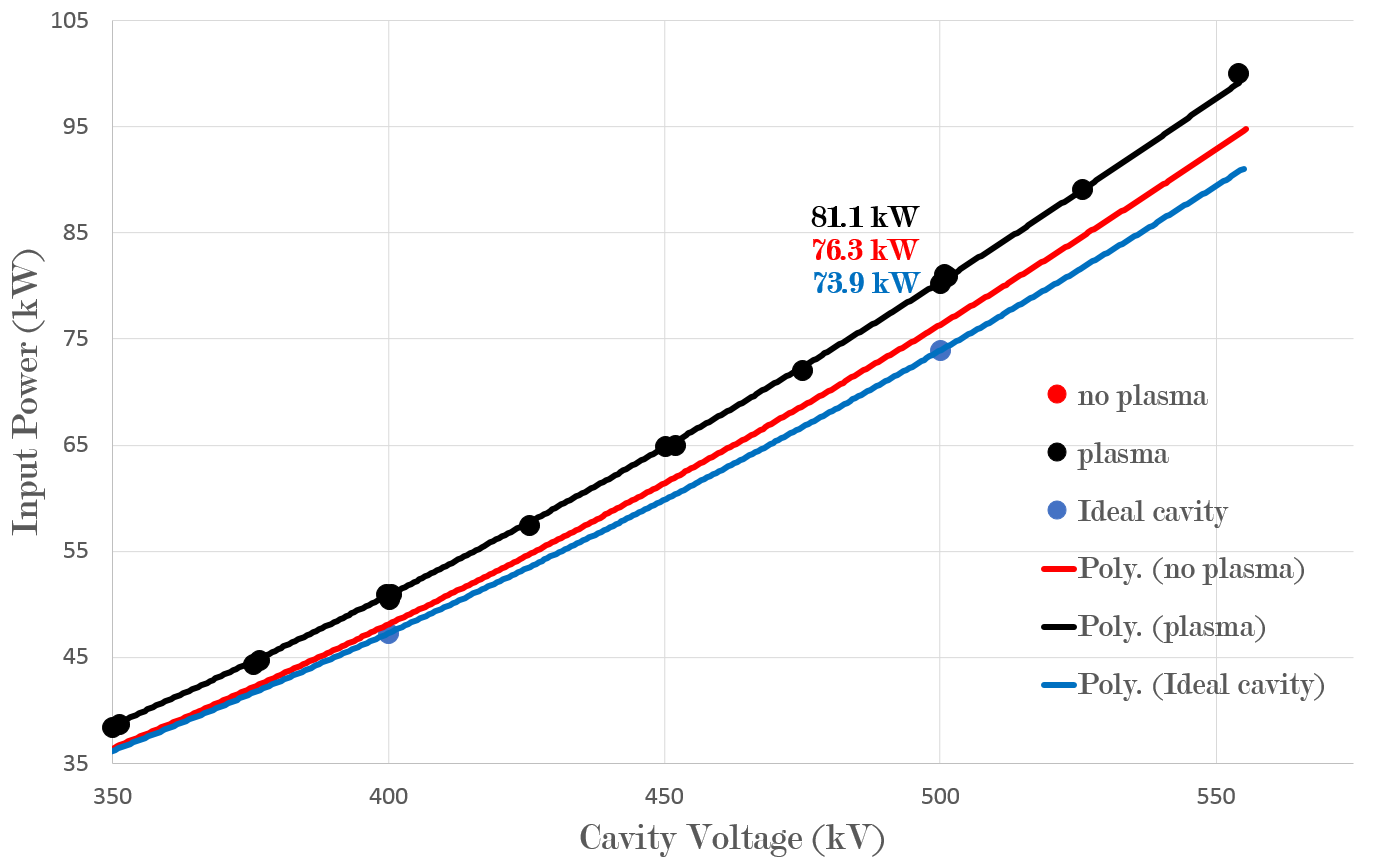}
  \caption{The power required by the existing Flat Top cavity (black) to reach a specified voltage is shown, as well as the ``ideal model" projected power requirements (blue). The plasma is ignited around 320 kV, and thus below this threshold several data points were taken. The data was used to create an extrapolation of the power needed in the cavity without the plasma ignited (red). Comparing the plasma vs no plasma data, it can be seen that at 500 kV the plasma pulls 5 kW of power out of the cavity. Comparing the no plasma scenario to the ideal model, one can see a 2.4 kW difference exists, which happens to match the leaked power seen in the ``realistic model".}
  \label{power}
\end{figure}

Lastly, one may have noticed that several of the cases studied, such as the ``realistic case" and 0.5 mm deviation, required less power to reach the 500 kV than the ``ideal case". This is due to an unforeseen reason, the particles are receiving an additional gain when traveling through the vacuum chamber. Figure~\ref{vacgain} shows the ``realistic case" model and the line integration used to determine the voltage gain in the cavity. The electric (top right) and the magnetic (bottom right) field patterns are shown as well. The integration through the vacuum chamber adds $\sim$0.5$\%$ to the voltage and $\sim$1.2$\%$ reduction on power needed to reach the field. Thus the leaking field is actually being utilized to accelerate the particles and thus the cavity requires less power to obtain the same gain. The gain, and its associated power reduction, depends heavily on the position of the lips. As the lips move the electric and magnetic fields move and rearrange in the vacuum cavity. As a result one would have to exactly know the position of the lips in order to take into account the vacuum chamber acceleration effect.

Figure \ref{power} shows the power inserted into the real life FTC and the corresponding voltage \cite{Plasma}. The comparison point between this study and the measured data is at 500~kV. The data shows that the cavity requires approximately $\sim$81 kW. This is 7.5 kW higher than the ideal case generated from the CAD drawings. As a result, it is already confirmed that an asymmetry exists in the cavity. The power was also compared to the realistic case, or the closest that could be achieved with the interferometry reading, Table \ref{table3}.

\begin{table}[h!]
 \centering
\hspace*{-0cm}

\begin{tabular}{ |P{1.5cm}|P{1.5cm}|P{1.5cm}|P{1.5cm}| }
 \hline
 \multicolumn{2}{|c|}{Power (kW)} &\multicolumn{2}{c|}{Leakage (kW)}  \\
 \hline
 Total &Leaked& SM7 Port &SM6 Port \\
 \hline
 73.6    &2.40&  0.3&2.10\\
 \hline
\end{tabular}
   \caption{This shows the total amount of power to reach 500 kV in the ``realistic model", as well as the associated leaked power.}
  \label{table3}
  \end{table}

The data shows, Table \ref{table2}, that the cavity needs approximately 71-72 kW to be retained in the cavity to reach 500~kV. However, this means in many cases more than 80 kW has to be inserted into the cavity due to the large leakage, suggesting the lips are moved by 1-2 mm. An important item to note is that in this study the lips are assumed to be straight the entire length of the cavity. However from the interferometry, data shows that the lips vary along their length. As a result, the only method to fully model the cavity would be to perform interferometry on the entire cavity as previously stated.

\begin{table}[h!]
 \centering
\hspace*{-0cm}

\begin{tabular}{ |P{1cm}||P{1cm}|P{1cm}|P{1cm}|P{1cm}||P{1cm}|P{1cm}|P{1cm}|P{1cm}|P{1cm}| }
 \hline
 \multicolumn{1}{|c||}{}&\multicolumn{2}{c|}{SM7 Top}&\multicolumn{2}{c||}{SM7 Bottom}&\multicolumn{2}{c|}{SM6 Top}&\multicolumn{2}{c|}{SM6 Bottom} \\
 \hline
 Shift (cm)& Total (kW)&Leaked (kW)& Total (kW)&Leaked (kW)& Total (kW)&Leaked (kW)& Total (kW)&Leaked (kW)\\
 \hline
 0.0		&73.9		&0.60		& 		& 		&&&&\\
 1.0		&72.3		&0.60		&72.2		&0.43		&72.3&0.61&72.2&0.45\\
 2.0		&72.6		&0.92		&72.4		&0.56		&72.7&0.97&72.5&0.64\\
 5.0		&74.6		&2.83		&74.0		&1.93		&74.9&3.19&74.4&2.39\\
 10		&81.0		&9.10		&79.5		&7.36		&82.4&10.7&81.2&9.17\\
 \hline

\end{tabular}

  \caption{The position of the cavity walls and the corresponding leakage out of the cavity. The leaked power is reported in kW, and the shifted step increments are in mm.}
  \label{table4}
  \end{table}

The walls of the cavity are also displaced in reality compared to the CAD drawings. However perturbation of the wall$'$s location has little influence on the cavity$'$s leaked power. The walls on each side were moved independently towards the center of the cavity to determine their effect with position changes of 1, 2, 5, and 10~mm. It can be seen from the data in Table \ref{table4} that the walls are approximately seven times less sensitive to motion as the lips. Therefore, the walls would have to be extremely deviated asymmetrically in order to produce the power leakage seen in the real cavity.

The power leaks out the end opposite from the side that was altered, which is exactly the opposite effect generated from the lips. Moving the walls inwards generates twice as much leakage as moving them outward the same distance. Moving both walls on the same side results in approximately the same leakage as moving one wall. Similar to the behavior of the lips, the bottom half is less sensitive to the walls location than the top half of the cavity. Setting the walls to the most accurate position measured (2 mm inward SM7, 1 mm outward SM6) results in the amount of leakage from moving a single wall 1 mm.

 \begin{figure}[h!]
 \centering
\hspace*{-0cm}
   \includegraphics[scale=.35]{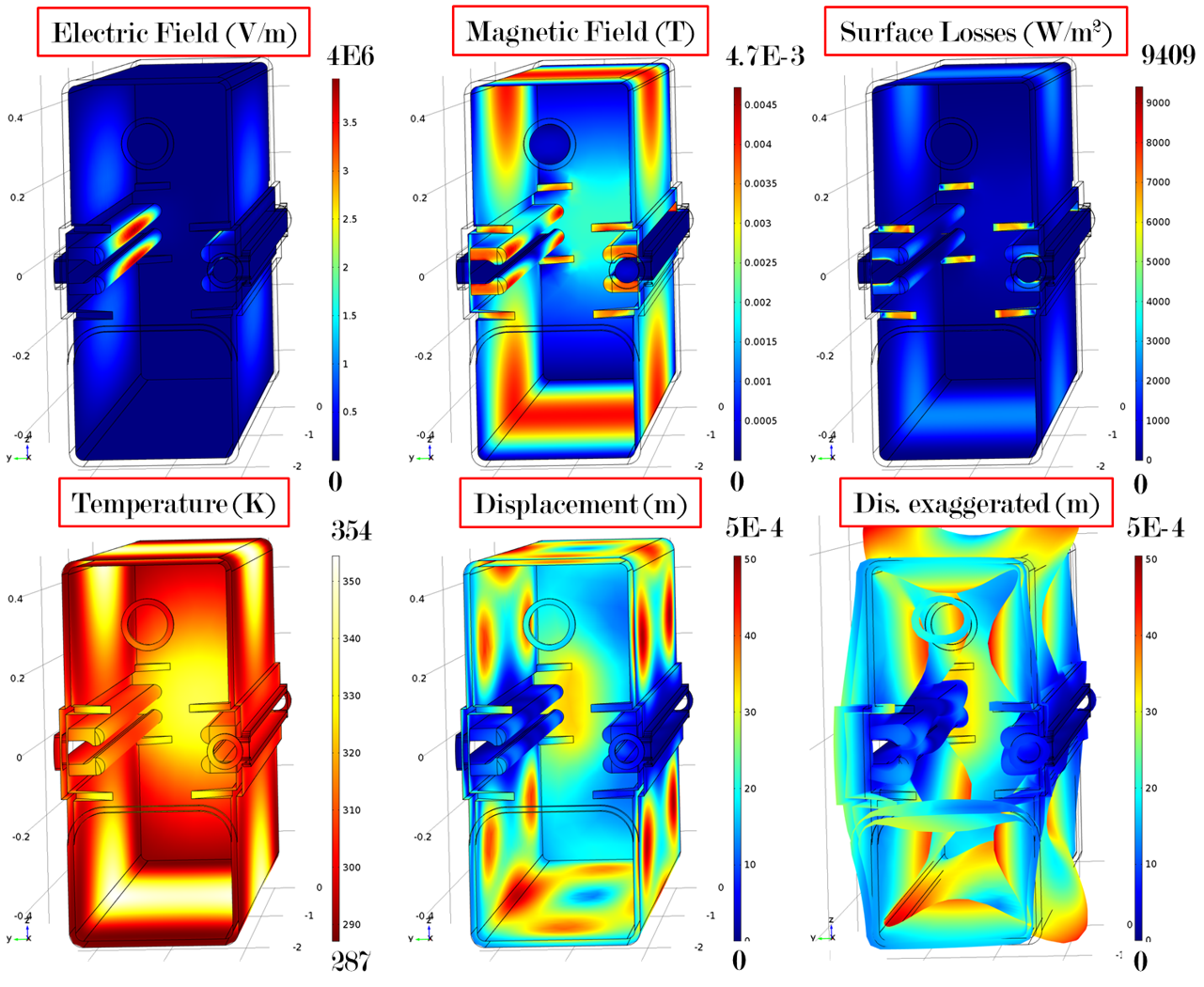}
  \caption{The FTC model created to determine displacement during operation. The model uses the electromagnetic fields to calculate losses and Lorentz forces on the cavity. The losses are then used to determine the heat produced in the cavity. The Lorentz forces, gravity, atmospheric pressure, and thermal expansion from heating are used to calculate the displacement.}
  \label{multi}
\end{figure}

A mechanical model, shown in Figure \ref{multi}, was created using the electromagnetic fields to determine heat deposition and Lorentz forces. The heat, gravity, Lorentz forces, and atmospheric pressure were then used to determine the deformation of the cavity. Cooling was applied to the simulation to reach stable equilibrium. The walls, not water cooled, had energy removed by convective cooling in an environment of 20$^{\circ}$C and the heat transfer coefficient of 5~W/(m$^{2}$~K). The surfaces with water channels were cooled using 300~W/(m$^{2}$~K), a conservative estimate. The cavity reached a steady state at 80$^{\circ}$C. The RF cavity has 20 temperature monitors placed around its exterior. The average measurement from the monitors near the peak temperature locations is 75.5$^{\circ}$C. Thus the model closely reflects the heating of the cavity with its cooling channels in operation.

The resulting cavity deformation showed that the largest deviation expected in the lips during operation was approximately 0.2~mm. The walls had a maximum deviation of 0.5~mm inward and 0.5~mm outwards. With this new information in hand, the models suggest approximately 720~W would be leaked out of the deformed cavity due to the heating and forces on the walls and lips. This is only 120~W more than the non-deformed case. Thus the inherent asymmetry in the cavity shape is the biggest contributor to the leaked power.

The model was completed in COMSOL Multiphysics$'$ -- RF Frequency, Heat Transfer, and Structural modules \cite{Comsol}. Once again tetrahedral mesh elements were used, but due to computational constraints, only 600,000~elements were utilized in the model. The thickness of the aluminum creating the shell was determined from the CAD model, and any anomalies generated using the "shelling" process were healed in CAD.

This study shows the level of tolerance needed to reduce the leaking power to a minimum. If the lips can be aligned within 1~mm, the amount of power leaked, and thus required to reach 500~kV, is reduced to a minimum. Similarly if the walls$'$ motion can be reduced to a few millimeters, then the power leaked is reduced to a minimum. The additional power saved could be utilized to increase the voltage to higher levels than currently capable.

Not all of the power leaked from the cavity is due to asymmetry of the cavity. A plasma is formed inside the vacuum space due to the leaked RF power. The plasma then proceeds to change the impedance of the system and thus pulls out additional power from the cavity. Figure~\ref{power} shows the power required by the cavity for the corresponding voltage measured. The top line (black) is the measured data from the cavity. However, the plasma was found only to ignite above 320~kV. Thus the data below this threshold allows us to create a projected power consumption for the cavity without the plasma. That line is shown as a red line. This illustrates that the plasma is coupling out nearly 5~kW more than when the plasma was extinguished. 

A second, and possibly more important finding is that only about 2.4~kW of power is needed due to the asymmetry of the cavity$'$s form. This is the difference between the extrapolated ``no plasma" line and the ``ideal case model". The reason that is of importance is that it matches the leaked power from the ``real model" as shown in Table \ref{table3}. This provides some evidence that the modeling performed is simulating reality accurately.

The result of these studies show that most likely the cavity$'$s lips are around 1 mm displaced. Similarly the walls are most likely within a 1~mm of symmetry. Future iterations of the cavity could call for more stringent tolerances during manufacturing and assembly. The other options is to develop methods to reduce the power capability to flow into the vacuum space.

\section{Possible Solutions}
The general tolerance of the cavity required is now known for future iterations of the FTC. However, the development and deployment of a new cavity will take several years. As a consequence, it would be highly desirable to identify a method of reducing the leaked power in the existing cavity until a new version can be installed. The first method conceived was to introduce a choke joint on the beam pipe just outside of the cavity boundaries.

 \begin{figure}[h]
 \centering
\hspace*{-0cm}
   \includegraphics[scale=.33]{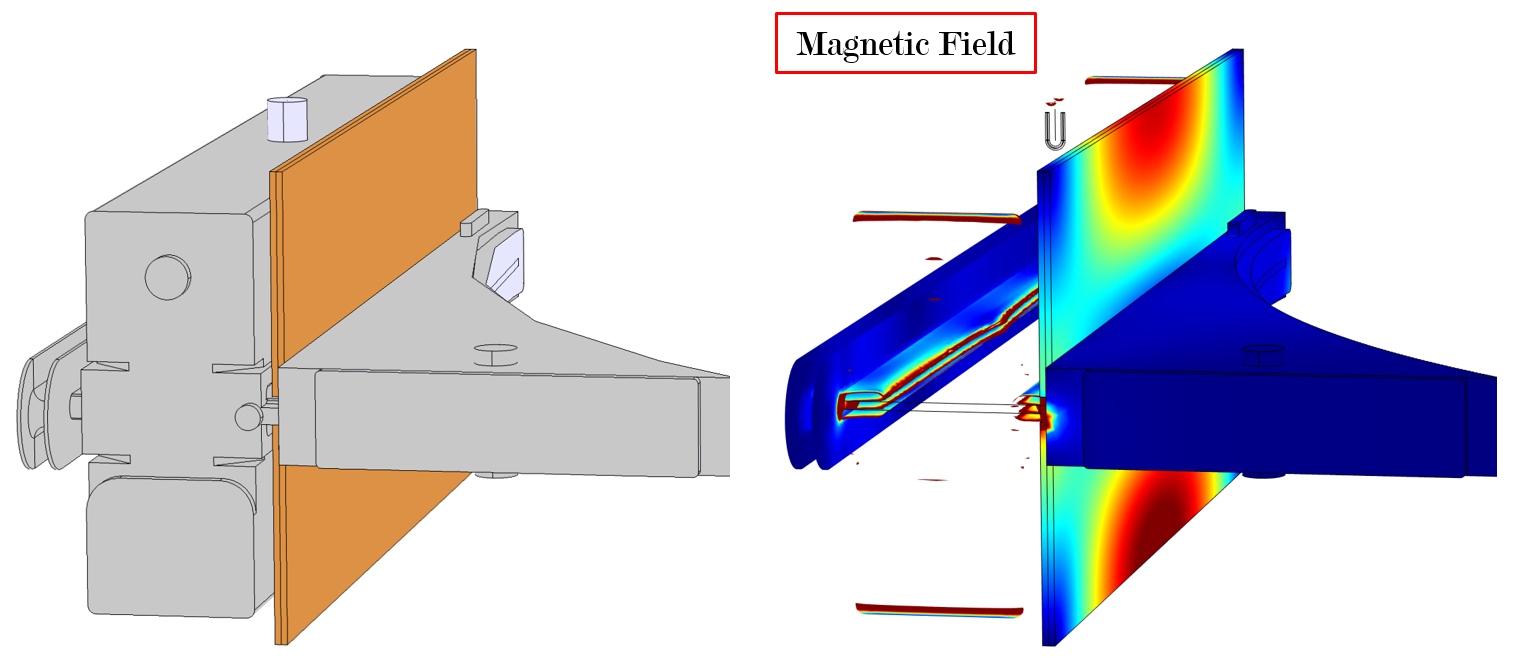}
  \caption{The choke joint geometry (shown in orange) was added to the previous model to reduce the fields leaked by $\sim$95$\%$. The bottom image shows the magnetic field within the choke that is approximately .2$\%$ of the maximum magnetic field value in the cavity.}
  \label{choke}
\end{figure}

Normally choke joints are used to create a low impedance bridge between waveguides. The joint introduces an extra path length of $\lambda$/2 to introduce the low impedance. In this particular situation we wish to create the exact opposite phenomenon. Thus a path length of $\lambda$/4 is introduced creating an extremely high impedance, thus dramatically reducing the power leaked out of the cavity. An image of the model, including the choke is shown in Figure \ref{choke}.

Introducing the joint in the ideal cavity causes a minimal reduction. The amount of power leaked on the SM7 side was reduced by 17\% and 9.5\% on the SM6 side. For the case where the top SM7 side lip was moved by 2 mm (worst case scenario), the leaked power was reduced from 11.9 kW to 571 W, or a 95\% reduction. This choke can therefore limit the leakage to nearly that of the ideal case if implemented correctly. The tolerance on the choke'$s$ height is within 3~cm. Thus if the height of the choke exceeds, or falls short of, the target height (56 cm from the beam plane), the power begins to leak substantially more.

Although the choke is quite successful in reducing the amount of power leaked out of the cavity, the geometry modeled is not immediately implementable in the real cavity. The choke would interfere with the cavity$'$s structural elements and hydraulic tuning systems. Therefore another method of reducing the leaking fields was modeled.

The distance between the lips vertically is 3~cm, however the distance grows quickly past the lips. A set of blocks were inserted into aperture bringing the effective height of the vertical beam gap to 3~cm. The ``shim" blocks effectively occupy volume seen by the fields in the cavity, and thus attenuates the fields traveling out of the cavity.

 \begin{figure}[h]
 \centering
\hspace*{-0cm}
   \includegraphics[scale=.33]{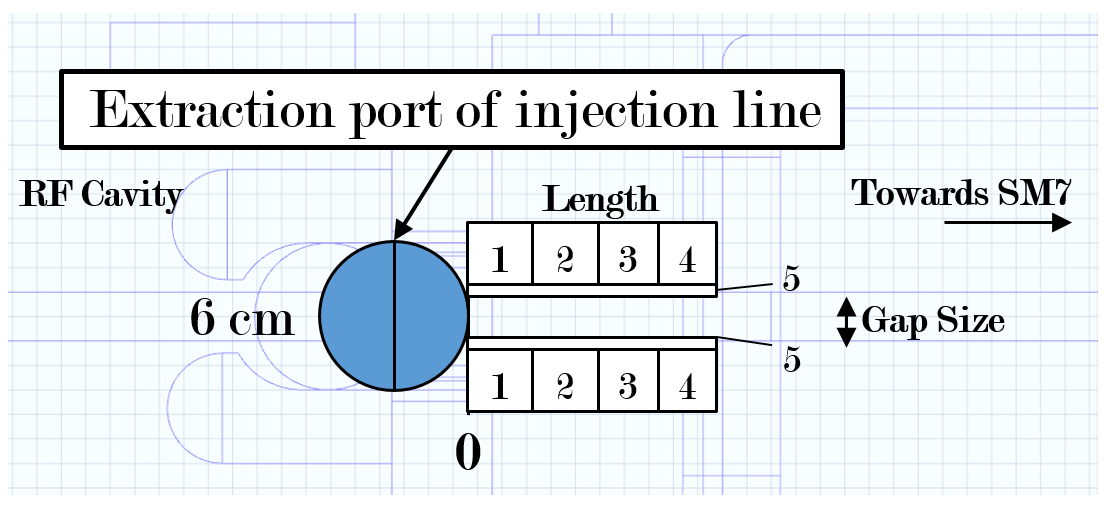}
  \caption{The setup for the insertion of the metal shim into the beam aperture. The shim fills the volume in the beam aperture to reduce the vertical height to 3~cm or the same height as the lip spacing (on left). The shims are extended to a maximum of 10.2~cm where the shim would begin to penetrate the vacuum cavity on the SM7 side. Lastly, the gap size was reduced to 2.5 cm, the size of the gap in the sector magnets (5).}
  \label{shim}
\end{figure}

The shims were only inserted on the SM7 side as it is thought most of the leaked power causing the plasma generation is through this port. The shims were inserted to line up with the injection line outlet port so as not to interfere with cyclotron operation. Starting with a 3~cm block, the size was extended up to 10.2~cm as shown in Figure \ref{shim}. The result of the shim insertion was a significant drop in leaked power.

\begin{table}[h!]
 \centering
\hspace*{-0cm}

\begin{tabular}{ |P{2cm}||P{1cm}|P{1cm}|P{1cm}| }
 \hline
 \multicolumn{1}{|c||}{Description}&\multicolumn{1}{c|}{Length}&\multicolumn{2}{c|}{Power at 500 kV} \\
 \hline
 Shift& (cm)&Total (kW)&Leaked (kW)\\
 \hline
 No Shim&-		&82.7		&11.9	\\
 \# 1		&2.9		&78.3		&7.6	\\
 \# 2		&5.4		&76.5		&6.0	\\
 \# 3	&7.9		&75.3		&4.9	\\
 \# 4	&10.2		&74.4		&4		\\
 \# 5	&10.2		&73.6		&3.3		\\
 \hline

\end{tabular}

  \caption{The table shows the length of the shim inserted into the cavity model and the corresponding reduction in leaked power. Without a shim, the power leaked is 11.9 kW. However with progressing length of shim (\# 1-4) the power can be reduced by 66\%. If the gap size is reduced to 2.5~cm (\# 5) the power leaked is reduced by 75\%.}
  \label{shimtable}
  \end{table}

Without the shim, the leaked power was 11.9~kW as illustrated in Table \ref{table2}. When the 3 cm blocks were inserted the power dropped to 7.6~kW as shown in Table \ref{shimtable}. The longest shim modeled, 10.2~cm, allowed only 4~kW of power to be leaked, or a 66\% reduction. If the vertical gap is reduced to 2.5~cm (\# 5) instead of 3~cm, the power drops to 3.3~kW. The 2.5~cm vertical gap size matches the size of the aperture in the sector magnet and is a possibility. The results of this study show that if the shims were inserted into the cavity$'$s aperture, the amount of power leaked could be reduced by 75\%. As to whether the plasma will be eradicated or moved closer to the cavity is unknown. If the plasma is moved closer, it would at least reduce the deposition of metal on the ceramic features in the electrostatic septa.

Introducing shims on the sides of the beam pipe to reduce the radial aperture had no effect on the power leakage. The shims were placed 1 cm off the injection and extraction radial position with respect to the center of the cyclotron. Since this minimal radial aperture, extraction radius minus injection radius, is still larger than one wavelength of the fundamental frequency of the cavity (150~MHz), there was no reduction in power leaked. Lastly, a simple geometrically symmetric coupler similar in size and shape with a semi-circle loop, shown in Figure \ref{choke}, replaced the real coupler in the cavity and showed that asymmetry was indeed reduced. When the symmetric coupler was inserted, the amount of leaked power dropped from 3.3~kW to 2.9~kW with no other alteration to the cavity. Thus in the future it would be beneficial to make the coupler symmetric. A summary of all the results is shown in Table \ref{shimtable}.
\section{Window Leakage}

The FTC has a total of twelve ports. Two of the ports are the beam pipes that are attached to the sector magnets and another two are entrance and exit of the injection beam line that travels through the FTC. Five of the other ports are capped by metal and all reside in the vacuum space between the cavity and SM7. The three remaining ports are capped with ceramic windows.

In the previous studies, the leakage through the beam apertures were studied in detail. The losses on the ports with metal caps were also studied, as they were considered part of the metal cavity structure. Thus only five ports remain whose power leakage has not been studied: the beam injection line entrance and exit, the window in the main cavity, and two windows located on the radially outward face (from the cyclotron$'$s center) of the vacuum chamber.

The two windows located in the outward face of the vacuum chamber are covered by a ceramic window made of radiation resistant Schott BK7G18 glass \cite{Schott}. The inner surface of the window is covered with a thin coating of Indium Tin Oxide (ITO) that is specified to be 20~$\Omega$/$\Box$. J. Eite and Spencer \cite{ITO} produced experimental data illustrating the attenuation of the power through the film as a function of frequency. The Flat Top operating at 150~MHz would have 1/1000th of the power incident on the window leaked out of the cavity. One can therefore treat these two windows as being covered with metal, which was already the condition applied in the previous studies.

Hence there are only three port remaining whose power leakage has not been studied: the injection line entrance and exit and the borosilicate window located in RF cavity itself. Due to the location of the injection line, it is expected that these will leak very little power. In contrast the borosilicate window is expected to leak substantially more power out of the cavity, but not nearly as much as the beam apertures as the field components and current flow are reduced in the window$'$s location.

\begin{table}[h!]
 \centering
\hspace*{-0cm}

\begin{tabular}{ |P{2cm}||P{1cm}|P{1cm}|P{1cm}|P{1cm}|P{1cm}|P{1cm}|P{1.25cm}| }
 \hline
 \multicolumn{1}{|c||}{}&\multicolumn{2}{c|}{Power (kW)}&\multicolumn{5}{c|}{Leakage (W)} \\
 \hline
 Case& Total&Leaked&SM7 Port& SM6 Port& Inj. Port&Ext. Port&Main Cavity\\
 \hline
 Ideal		&74.6		&1.33		&347&276&71&23&607\\
 Real		&74.3		&3.00		&265		&2082		&71&23&608\\
 SM7 Top 2mm		&83.4		&13.0		&11.9		&385		&70&23&639\\
 \hline

\end{tabular}

  \caption{The RF window in the main cavity leaks approximately 600 W no matter the asymmetry introduced in the cavity. It also shows that almost no power leaks out of the injection and extraction ports.}
  \label{window}
  \end{table}

The power flow was simulated out of these ports and the results for three different cases are shown in Table \ref{window}. The leakage out of the beam line ports are minuscule compared to the input power. The borosilicate window, on the other hand, leaks approximately 600~W. This study also shows that RF window and injection line ports are basically insensitive to the asymmetry in the cavity. The main cavity window should have almost no dependence on the asymmetry of the cavity as the RF fields at the stored energy to produce 500~kV is uniform across all models. To improve power efficiency the window could be covered with an aluminum cap or coat the interior face with ITO to negate the power leakage.

\section{Conclusions}

The studies performed above were to analyze the existing RF cavity and have readily available an electromagnetic model of the existing cavity and its properties in case the FTC were to have a serious failure. The data here shows that the cavity lips are of critical importance in reducing the amount of power lost. Keeping the lips within a millimeter tolerance can potentially reduce a substantial amount of power lost in this flavor of cavity in future designs.

There are three items that could be put into action with the existing cavity to reduce losses. The first is to cover the borosilicate window in the main body of the cavity with either an aluminum cover or coat the inner face with ITO. This could reduce power consumption by 600~W at 500~kV. A more dramatic and invasive method would be to introduce the shims into the cavity. Though not as effective as the choke joint, the insertion of 10.2~cm long blocks that squeeze the vertical aperture to 2.5 - 3~cm can reduce the leakage by as much as 75\%. The effect on the plasma phenomenon is unknown, but reducing the leaked power is better for the cyclotrons stability and sustainability. The third option is to install an asymmetric tuning system to the exterior of the FTC where each of the four quarters would have independent control of pressure applied. This would allow compensation for distortion of any of the lips over time.

The studies performed here have identified paths to reduce the amount of leaked power entering the vacuum space in the FTC. Though this cavity is unique, the phenomenon investigated here will occur in other high power cyclotrons. As a result of these studies, the design of a FTC upgrade has been manipulated to avoid these issues. 

\section{Acknowledgments}
The research leading to these results has received funding from the European Community's Seventh Framework Programme (FP7/2007-2013) under grant agreement n.$^{\circ}$290605 (PSI-FELLOW/COFUND). The author would also like to acknowledge Tino Howler for the interferometry data, Harold Siebold for assistance with CAD drawings, and Markus Schneider for assistance in RF measurements. Nathaniel Pogue also holds Visiting Scientist status at Texas A\&M – collaborating on High Intensity Cyclotrons.

\section{Bibliography}
\bibliography{els}
%
%
%
%

\end{document}